\documentclass[twocolumn,aps,prb,raggedbottom,nobalancelastpage,amssymb,superscriptaddress,showpacs]{revtex4-1}

\usepackage{amsmath}
\usepackage{amssymb}
\usepackage{amsfonts}
\usepackage{dsfont}
\usepackage{graphicx}
\usepackage{bm}
\usepackage{color}[]
\usepackage{appendix}
\usepackage{epsfig}
\usepackage{empheq}

\usepackage{soul}
\setstcolor{red}
\sethlcolor{yellow}

\newcommand\ket[1]{\left|#1\right\rangle}

\newcommand\avg[1]{\left\langle#1\right\rangle}
\newcommand\Avg[1]{\langle#1\rangle}

\newcommand{\wv}[3]{{}_{#1}\Avg{#2}_{#3}}

\newcommand\eq[1]{Eq.~(\ref{#1})}
\newcommand\eqs[2]{Eq.~(\ref{#1}, \ref{#2})}
\newcommand\fig[1]{Fig.~\ref{#1}}

\begin{document}
\title{Weak measurement of cotunneling time}

\author{Alessandro Romito}
\affiliation{\mbox{Dahlem Center for Complex Quantum Systems and Fachbereich Physik, Freie Universit\"at Berlin, 14195 Berlin, Germany}}
\author{Yuval Gefen}
\affiliation{Department of Condensed Matter Physics, Weizmann
    Institute of Science, Rehovot, Israel}

\begin{abstract}
Quantum mechanics allows the existence of ``virtual states'' that have no classical analogue.
 Such “virtual states
defy direct observation through strong measurement, which 
would destroy the volatile virtual state.  
Here we show how a virtual state of an interacting many-body system
can be detected employing a  weak measurement protocol with postselection.
We employ this protocol for the
measurement of the time it takes an electron to tunnel through a
virtual state of a quantum dot (cotunneling).
Contrary to classical intuition, this
cotunneling time is independent of the strength of
the dot-lead coupling and may deviate from that predicted by
time-energy uncertainty relation. Our approach, amenable to
experimental verification, may elucidate an important facet of quantum
mechanics which hitherto was  not accessible by direct
measurements. 
\end{abstract}

\pacs{
}

\maketitle

\section{Introduction} 
\label{introduzione}
An important aspect of quantum mechanics is the existence of states
that have no classical analogue.
Such ``virtual states'' cannot exist in classical physics as they violate
energy conservation. 
It is commonly suggested that their presence within quantum mechanics
as short-lived states is allowed by the uncertainty
principle~\cite{Messiah1999} $\Delta t \, \Delta E \sim \hbar$. 
These states, being volatile, are destroyed by a strong measurement,
and are therefore  inaccessible to direct detection.
By contrast, weak measurement, along with its  weak backaction, may
provide us with a nondestructive probe into  virtual states. 
In fact, weak measurement based protocols with postselection
[\emph{weak values} (WV)]~\cite{Aharonov1988} have been employed with remarkable success in explaining quantum paradoxes ~\cite{Aharonov2005}, detecting
and amplifying  weak signals~\cite{Hosten2008,Dixon2009}, directly
measuring a wave-function~\cite{Lundeen2011} and devising protocols
for quantum states discrimination~\cite{Zilberberg2013}.

The primary goal  of this paper is to extend the utility of WVs to
the arena of many-body states, specifically  probing virtual many-body
states. 
This task is accomplished here for the first time, by specifically considering
the process of cotunneling~\cite{Averin1990,Glazman2005}.
In the latter,  electrons are transported between a source ($S$) and a drain ($D$) through a quantum dot (QD); the QD is tuned such that the addition of an extra charge to it is classically forbidden (the Coulomb blockade regime)~\cite{Aleiner2002}. 
Nevertheless, an electron can enter the QD and later exit, forming a
(short-lived) virtual many-body correlated state. 
This cotunneling process is qualitatively different from a single
particle tunneling under the barrier.
We design a weak value protocol, amenable to experimental test,
tailored to measure the lifetime of such a many-body virtual state.  
We anticipate that our demonstration of feasibility of such a protocol
will pave the road to the study of a host of many body problems that
involve virtual states.  

The second goal of our work is the study of the specifics of
cotunneling life-time.  
The lifetime of a virtual state associated with the tunneling of a single
particle under a barrier has been studied  extensively with a variety
of approaches~\cite{Condon1931,Wigner1955,Buttiker1982,  Buttiker1983,Mullen1989,Sokolovski1987,Sokolovski1993,Steinberg1995,Choi2013}. 
There are several time scales involved in this process: the dwell
time, $\tau_\textrm{dwell}$, marks the lifetime of the virtual state
regardless of whether the  electron is eventually transmitted (to $D$)
or reflected (to $S$); the traversal time is the lapse between the
disappearance of an electron from $S$ and its appearance in  $D$. 
For the tunneling of a single particle it has been
shown~\cite{Buttiker1982} that the time is related to the imaginary
velocity of the particle under the barrier. 
Determining the  traversal time in the many-body cotunneling case
poses a more difficult challenge, elucidated below.

Here we show that naive  expectations based on
analogy with a single particle tunneling are unfounded. 
Strikingly we also find that
the cotunneling time, $\tau_{\textrm{cot}}$,  may not be related to
the time-energy uncertainty relation~\cite{Mandelstam1945} $\Delta E
\, 
\tau_\textrm{cot} \sim \hbar$, where $\Delta E$ represents the violation
of energy conservation in the virtual state. 
Finally we note that, while by classical intuition the transmission
time through a QD should depend on the dot-lead tunneling matrix
element, this turns out not to be the case here. Our results are
summarized in Table~1. We find that $\tau_{\textrm{cot}}$ depends
parametrically on whether the cotunneling is dominated by elastic or
inelastic processes.

In the following, after defining our model of a detector weakly coupled
to a quantum dot tuned to the cotunneling regime, we review the
measurement time in the regime of sequential tunneling, where the
transport occurs through classical probability rates of tunneling in
and out of the dot.
 Employing the same setup, we present a
semi-heuristic  procedure through which we define the cotunneling
time. We then show that this definition coincides with  the quantity
obtained relying on  a weak-value based protocol.
 We finally compute
the cotunneling time in the various relevant parameter regimes.

\begin{figure*}[ht]
\centerline{\includegraphics[width=.85\textwidth]{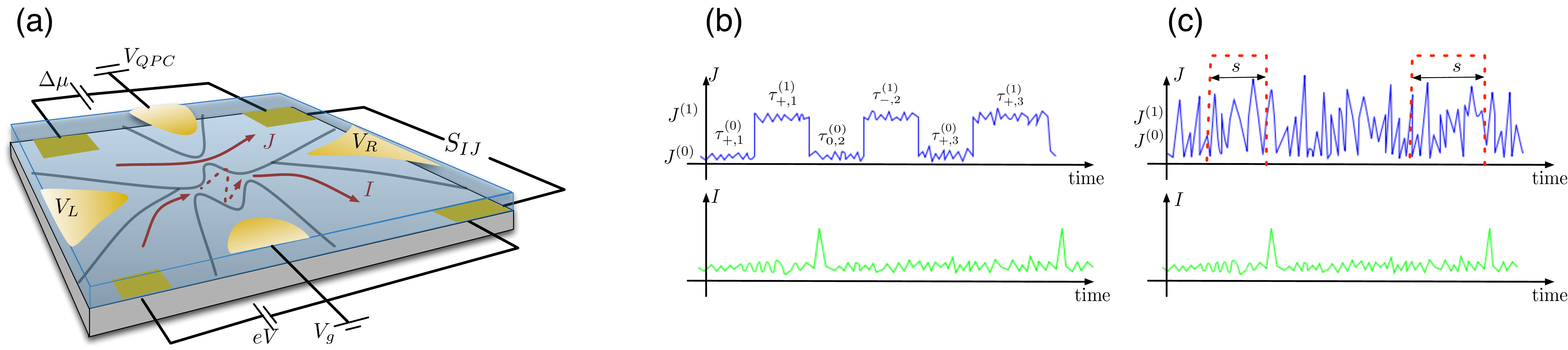}}
\caption{A measurement scheme of the cotunneling time.
(a) Sketch of a quantum dot coupled to a
QPC detector. The transmission through the QPC is 
affected by the presence of an extra electron in the QD.
The top gates $V_L$ and $V_R$  control the tunneling rate to the dot,
$V_{\textrm{QPC}}$ the unperturbed transmission  through the QPC, and $V_g$ the
charging energy in the dot.  
$V_g$ allows us to tune the system from the sequential tunneling
to  the cotunneling regime.
The transport through the QPC and the QD is controlled by
the voltage bias $\Delta \mu$ and $eV$ respectively.
The current-current correlation $S_{IJ}$ is sensitive to the 
excess number of electrons, $N$, in the dot.
(b) Typical detector signal and current through the dot in the
sequential tunneling regime, performing strong measurement with the
QPC.  
The time an electron spends in the dot
can be classified according to the occurrence of a subsequent positive
current pulse in the drain ($\tau_{+,i}$), or the absence thereof ---back-reflection ($\tau_{-,i}$). 
The average sequential  tunneling time can be directly obtain by
averaging over the durations, $\{ t_{+,i}^{(1)} \}$, of the relevant
QPC signals.
(c) Typical detector signal and current through the dot in the
weak measurement regime.
Straightforward classification of events as in (b) is not
possible. 
The signal of the QPC preceding a pulse of current through the quantum
dot,  has to be weighted by an appropriate weight function. 
The latter should account for the QPC current signal, which  precedes
and is  directly related to the electron detected at the QD's drain terminal. 
In the cotunneling regime, the interval between successive tunneling
events is longer than the cotunneling time, hence one may discard 
the weighting function.
}\label{fig:sistema}
\end{figure*}

\section{Model and setup}
\label{modello}
Our setup [cf.\ Fig.~\ref{fig:sistema}(a)], consists of a system (a
quantum dot weakly coupled to leads) and a detector [a quantum point
contact (QPC)].  
The detector measures the system through the electrodynamic coupling
between them. 
The way this setup is defined it is suitable to discuss both transport
that involves real processes (sequential tunneling) as well as virtual
processes (cotunneling). 
The relevance of many-body states is self-evident here. 
The corresponding Hamiltonian is given by
\begin{align}
\label{eq:Hamiltoniana}
H=H_0+H_T+H_{\textrm{int}}+\mathcal{H}_{\textrm{detector}} ,
\end{align}
where $H_0$ represents the isolated (but voltage biased) QD and the
uncoupled source ($S$) and drain ($D$) leads; $H_T$ stands for the
dot-leads tunnel coupling and $H_{\textrm{detector}} +
H_{\textrm{int}}$ describe the detector dynamics and its interaction
with the QD.
Specifically, the part of the Hamiltonian concerning the system (the
QD and the leads) consists of the following terms: 
\begin{align}
\label{h0}
H_0=\sum_{\alpha=S,D} \sum_{k} \epsilon_{\alpha,k}
c^\dag_{\alpha,k}c_{\alpha,k}+ \sum_h \epsilon_{0,h} d^{\dag}_h d_h +U.
\end{align}
It describes the isolated QD, and the
uncoupled source and drain leads.
The single particle terms for the source ($S$) and drain ($D$) leads
and the QD are expressed in terms of the fermionic field
operators, $c_{S,k}$,  $c_{D,k}$, $d_h$.
The charging energy contribution~\cite{Aleiner2002}
\begin{align} 
U = E_C (N_S+N_D-N_g)^2
- \frac{eV}{C_{\Sigma}} (C_S N_S - C_D N_D),  
\label{eq:ancora}
\end{align}
with $N=\sum_\alpha d^{\dag}_\alpha d_\alpha$,
characterized
by the energy $E_C$, depends on the charge entering the dot from the
left (right) lead, $N_{S(D)}$. $N=N_S+N_D$ is the extra charge on the dot.
\eq{eq:ancora}
provides the explicit dependence of the charging energy $U$
on dot-source,  $C_S$, dot-drain $C_D$, and dot-gate $C_g$ capacitances  ($C_{\Sigma} = C_S+C_D+C_g$ is the total capacitance), as well
as on the source-drain voltage bias, $eV$, and on the gate voltage, through $N_g
=V_g/(e C_g)$.

The leads-QD tunneling operator
$H_T= \sum_{\alpha=S,D} T_{\alpha} + \textrm{h.c.}$, with  $T_{\alpha}=
\sum_{k,h} \gamma^{(\alpha)}_{k,h} d_h^{\dag} c_{\alpha,k}$, are written
in terms of source and drain tunneling amplitudes $\gamma^{(\alpha)}_{h,k}$.
The current operator in the QD is
\begin{align}
\label{corrente}
I=\partial_t\sum_kc^\dag_{D,k}
c_{D,k} =i (T_D-T_D^\dag),
\end{align}Transport via virtual processes is $\propto \gamma^4$, and can be
classified into inelastic and elastic cotunneling, corresponding to
the state of the QD being modified or unmodified respectively,
following a tunneling event.   
In any case the virtual occupation of
the dot involves a many-electron correlated state.

The detection of the excess charge on the dot, $e N$, is carried out by
a quantum point contact (QPC) capacitively coupled to the dot, which is
routinely employed in experiments as a charge
sensor~\cite{Field1993,DiCarlo2004,Sukhorukov2007}.
The QPC is modeled as a scattering potential for impinging electrons
through
\begin{align}
\label{point-contact}
H_{\textrm{detector}} \equiv  H_{\textrm{QPC}}= \sum_{i=l,r}\sum_k v \, k \, a_{i,k}^{\dag}a_{i,k},
\end{align}
 with $a_{i,k}$ being the annihilation operator for the left-
($i=l$) and right- ($i=r$) moving scattering states; here $v$ is the
magnitude of the electron velocity.
The addition of an extra electron to the QD ($\Avg{N}=1$) results in
the modification of the QPC scattering potential.
The coupling to the QD is specified by $H_{\textrm{int}} = N
X$~\cite{Aleiner1997,Korotkov2001} with
\begin{align}
\label{interazione}
X=\frac{v(\delta \kappa +i u)}{L\sqrt{\kappa (1-\kappa)}}\sum_{k,h} a^\dagger_{r,k}a_{l,h} +\textrm{h.c.}.
\end{align}
It describes a general complex
back-scattering amplitude due to the modification of the scattering
potential due to the occupation of the QD by an extra electron.
The QPC has a chemical potential bias, $\Delta_\mu$, which defines the
detector's bandwidth and is assumed to be the largest energy scale in
the problem. The operator associated with the QPC signal is then the
current through the QPC~\cite{Korotkov2001,Blanter2000},
\begin{align}
J= \frac{ev}{L} \sum_{k,p} \left[ \sum_{i=l,r} \kappa (-1)^{i} a_{i,k}^\dag
a_{i,p} \right.& &  \nonumber\\
\left. \phantom{\sum_{k,p}} + \sqrt{\kappa (1-\kappa)} (i e^{-i (k-p) x} a_{r,k}^\dag a_{l,p}
+\textrm{h.c.}) \right],&& 
\end{align} 
Here $\kappa$ is the transmission probability through the QPC, $L$ is
the QPC length, and $x$ plays the role of a regularization parameter. 

\paragraph*{Various scales of charging energy.}
In view of the calculations of the cotunneling current and cotunneling
time, we conveniently denote by
$\Avg{U(T_S)}$ the change of charging  energy due to the tunneling of
an extra  charge from the  source into
the dot, and by $\Avg{U(T_S^\dag)}$ the corresponding change due to
the exit of a charge to the source. 
With the equivalent notation of the tunneling to/from the drain ($D$), we
can generally denote the modification of the charging energy due to a
certain sequence of tunneling events as  $\Avg{U(T_\alpha , \dots,
  T_\beta)}$ with $\alpha,\beta =S,D$.
In the Coulomb blockade regime where the charging energy $E_c$ is the
largest energy scale of the quantum dot's dynamics, we consider only
the states with $N=0,1$ excess electrons in the dot.
Then we obtain the relevant charging energies in the cotunneling processes directly
from \eq{eq:ancora}: 
\begin{subequations}
\label{eq:energia-carica}
\begin{align}
& \Avg{U(T_S)} \equiv E_{+l} ,\,\, \Avg{U(T_D^\dag)} \equiv E_{-r}  , \\
& \Avg{U(T_D)} = E_{+l}+eV , \,\, \Avg{U(T_S^\dag)} = E_{-r}+eV, 
\\
& \Avg{U(T_S T_D^\dag)} = -eV , \,\,  \Avg{U(T_S^\dag
  T_D)} = eV ,  \\
& \Avg{U(T_S T_S T_D^\dag)} = -eV +E_{+l} ,\\
& \Avg{U(T_D T_D  T_s^\dag)} = 2 eV +E_{+l}  , \\
 & \Avg{U(T_S^\dag T_S^\dag T_D)} =
2 eV +E_{-r}  ,\\
& \Avg{U(T_D^\dag T_D^\dag T_S)} = -eV +E_{-r} ,
\end{align}
\end{subequations}
Below we will focus on the
limit $eV \ll E_{-r} \ll E_{+l}$, in which cotunneling is dominated by
particle-like processes rather than hole-like processes; hereafter we simply
set $E_{+l} \equiv E_C$.

\section{A heuristic approach}
Before addressing the cotunneling regime, let us discuss how the
detection scheme of Fig.~\ref{fig:sistema} works in the sequential
tunneling regime,  which is a real (non-virtual) process.  
Referring to   Fig.~\ref{fig:sistema}(b), we note that the entry of the
$i$-th electron into the QD may result in a successful (unsuccessful)
sojourn, $\tau_{+,i}^{(1)}$ ($\tau_{-,i}^{(1)}$), at the end of which
the electron is transmitted to the drain (is backscattered). 
The current through the QPC, $J$, is a two-valued signal, where the
two values, $J^{(0)}$ and $J^{(1)}$, are associated with the absence
or presence of an extra electron on the QD. 
Noting that a peak in the current through the QD, $I$, signals a
successful tunneling event (we neglect processes  where the electrons
hop from the drain to the dot), one can easily extract the time of
sequential tunneling, $\Avg{\tau^{(1)+}}_\textrm{seq}$ from
\begin{align}
\label{eq:tunneling-sequenziale}
\Avg{J_+} = (J^{(1)}-J^{(0)}) \Avg{\tau^{(1)}_+}_\textrm{seq} \Avg{I}/e ,
\end{align}
which defines the tunneling time in terms of the excess current in
  the QPC conditional to the occurrence  of a  tunneling event
  through the QD,  $\Avg{J_+}$, and the average tunneling current $\Avg{I}$. A
  rate equation analysis reveals  that $<\tau^{(1)}_+>_\textrm{seq}$
  depends on the source and drain tunneling rates, $\Gamma_S$ and
  $\Gamma_D$ respectively,  yielding $\Avg{\tau^{(1)}_+}_\textrm{seq}=
  1/(\Gamma_S+\Gamma_D)$.

While \eq{eq:tunneling-sequenziale} is straightforwardly applicable to
experiments with strong QD-detector coupling~\cite{Sukhorukov2007},
$\Avg{J_+}$ cannot be directly addressed in the weak measurement regime.
The signal $J^{(1)}-J^{(0)}$ is then masked by quantum noise
 [cf.\ Fig.~\ref{fig:sistema}(c)] ; it is
not  possible to uniquely determine the duration of each interval
$\tau^{(1)}_{+.i}$. 
This hurdle can be overcome  by introducing a (Poissonian) probability
distribution, $p(t)$, for the time intervals $\{ \tau^{(1)}_{+,i}
\}$. 
Evidently $p(t)$ depends on $\Avg{\tau_+^{(1)}}_{\textrm{seq}}$. 
The sequential tunneling time may be obtained through an average over such a
distribution, as 
\begin{align}
	\label{eq:tunneling-debole}
\Avg{\tau^{(1)}_+}_{\textrm{seq}} = \lim_{T \to \infty} \frac{\int_0^{T}  dt \, \int_0^t
ds\, P(s) [J(t-s) -J^{(0)}] I(t)}{T \,\Avg{ I } \,  (J^{(1)}-J^{(0)})} , 
\end{align}
where $P(t) = 1 - \int_0^t ds \, p(s)$ is the probability the
electron, entering the dot at $t=0$ remains in the dot at time $t$.
This is a self-consistent equation for
$\Avg{\tau^{(1)}_+}_{\textrm{seq}}$. 
A direct calculation shows that \eq{eq:tunneling-debole}
leads to the same results as \eq{eq:tunneling-sequenziale} (cf.\ Appendix~\ref{sequenziale}).

We now consider the case of cotunneling. Here we generalize
\eq{eq:tunneling-debole}  employing quantum mechanical current-current
correlations.  
We stipulate that these correlations decay in time faster
than the time interval between two consecutive cotunneling events, hence
we may neglect the cut-off due to the $p(s)$, and replace $J^{(0)}$ by
the average $\Avg{J}$. 
This relates the cotunneling time to the current-current correlation
function  $S_{IJ} \equiv \int_0^{\infty} ds \, \langle  I(t)  [J(t-s)
- \Avg{J} ] \rangle$  through 
\begin{align}
\Avg{\tau^{(1)}_+}_{\textrm{cot}} = \frac{
  S_{IJ}}{\Avg{I} (J^{(1)}-\Avg{J} ) } \,  .
\label{eq:cotunneling}
\end{align}
It is worth noting that the integration in \eq{eq:cotunneling} is only
over positive times. 

Evidently, to evaluate the time obtained in the cotunneling regime, a
microscopic treatment of the problem is due. 
For this we employ the Hamiltonian in \eq{eq:Hamiltoniana}), and
evaluate the averages in the Keldysh
formalism~\cite{Kamenev2005,Keldysh1964}. 
to this goal a time ordering of operators, $\mathcal{T}_K$ is introduced on a time contour
consisting of two branches corresponding to forward- or
backward-in-time parts of the contour. 
The operators are labeled by a subscript $+,-$ depending on whether
they belong to the former or the latter branch of the contour. 
To first order in perturbation in $H_{\textrm{int}}$, the correlator
$S_{JI}$ reads
\begin{align}
\label{SS}
S_{JI} = & -i \int_0^{\infty} d \tau  \int_{-\infty}^{\infty} d s \left[
\Avg{\mathcal{T}_K [I_-(t+\tau) N_+(s)]} g_{++}(s-t)
\right.  \nonumber\\
& \left.  -  \Avg{\mathcal{T}_K [I_-(t+\tau) N_-(s)]} g_{-+}(s-t) \right] ,
\end{align}
where $g_{++}(t-t')
\equiv \Avg{ \mathcal{T}_K [J_+(t') X_+(t)]}$, $g_{-+}(t-t')
\equiv \Avg{\mathcal{T}_K [ J_-(t') X_+(t) ] }$.
In the limit considered here,  where  $\Delta \mu $ is
the largest energy scale in the problem,  one obtains
$g_{-+}(t-t') = g_{++} (t-t') = (e \Delta \mu)/(2 \pi) (i \delta \kappa + u) \delta
(t-t'+\eta)$, and finally
\begin{align}
\label{eq:misura-QPC}
\Avg{\tau_+^{(1)}}_{\textrm{cot}} = \mathrm{Re}\left\{  \tau_{WV}
\right\} - (u /\delta \kappa) \, \mathrm{Im}\left\{\tau_{WV}
  \right\} ,
\end{align}
where $\tau_\textrm{WV}$ is an intrinsic quantity of the system with
the dimensions of time,
\begin{align}
\tau_{\textrm{WV}}  = \frac{\int_0^{\infty} \Avg{I(t) \left[ N(t-s)
      -\Avg{N} \right] }}{\Avg{I}} .
\label{eq:definizione}
\end{align}
In fact $\tau_\textrm{WV}$ is the complex time obtained by a direct
application of a weak value protocol to the cotunneling time.

\section{Cotunneling time from the weak value formalism}
\label{detector-ideale}

In this section we present the result for the cotunneling time as
obtained through a direct application of the weak value formalism.
In complete analogy with the problem of single particle tunneling time~\cite{Steinberg1995}, $\tau_\textrm{WV}$ is
obtained with the aid of an ideal detector whose dynamics is trivial
($H_{\textrm{detector}}=0$).

The result is obtained employing a simple model in which the
detector is modeled as a pointer coupled via $H=\lambda  \hat{p} N$, $N$ being
 the excess charge in quantum dot (measured in units of the electron
 charge $e$; $N$ may assume the values $+1$ or $0$),  and $q$ the position of the detector pointer
 (initially at $\Avg{q}=0$) with $[q,p]=i \hbar$. 
The detector is assumed to have no internal dynamics ($H_\textrm{det}=0$).
Measuring $\hat{q}$ at a time $\Delta t$ leads to $\Avg{q} = \lambda
\int^{\Delta t} ds \, N(s)$. 
One can \emph{interpret} this expression to obtain either (i) the time
averaged charge in the dot, $ e \Avg{N}= e\Avg{q(t)}/\lambda \Delta t$, where
$\Delta t$ is the duration of the measurement, or (ii) the average
time the particle spends in the dot, $\tau = \avg{q}/\lambda $. 
In the latter interpretation it is important that the charge exists in
quantized units of $e$, and that during the measurement time $\Delta
t$ at most a single cotunneling event takes place.
In the case of sequential tunneling, this procedure results in
$\Avg{q}/\lambda$ being exactly the {\it dwell time} (as distinct from
the cotunneling time) of the particle in the QD. 
We assume this is a valid measurement of the dwell time also in the
regime of cotunneling. 

In order to address the time the particle spends in the dot
conditional to a later successful cotunneling event (which takes the particle to
the drain), we can make use of the weak value formalism~\cite{Aharonov1988}.
The signal in the detector, conditional to a successful cotunneling
through the QD, is expressed as $\tau_{\textrm{cot}} =
\wv{f}{q(t)}{0} / \lambda = \mathrm{Re} \{
\wv{f}{\tau_\textrm{WV}}{0}\} $, where $f$ indicates that 
the average has to be taken  conditional to the
postselection of a certain state $\ket{f}$ of the system. 
In the weak measurement regime,  $\tau_{\textrm{WV}}$ is the weak
value of the measured observable, hence 
\begin{equation}
\label{eq:espressione-semplice}
\tau_{\textrm{WV}} = \int^t\wv{f}{N(s)}{0}  =  \frac{\int^{t} ds \, \Avg{\Pi_f (t)
     N(s)}}{\Avg{\Pi_f(t)}}  ,
\end{equation}
where $\Pi_f$ is the projection into the postselected state.

In order to specify the postselection of the cotunneling process we
consider a simple picture where an electron, initially
in the source reservoir, can eventually reach the drain. 
The correlations between subsequently impinging electrons are neglected,
as well as the virtual occupation of the dot by processes originating
from the drain.

The projector onto the postselected state (i.e., successful cotunneling) at a time $\Delta t$ is
$N_D(\Delta t)$ (where the excess particle number, $N_D$, is measured
from the reference value before the cotunneling process started taking place). 
The postselection is, in fact, the result of a continuous measurement
over the interval $\Delta t$, which accounts for all possible arrival
times of the electron in the drain during the time interval $[0,\Delta
t]$.
This can be properly taken into account by summing the probability of
tunneling at any time and noting that it can be expressed via the
current operator at the drain as $N_D(\Delta t) = \int^{\Delta t}
dt \, I(t)$.
We therefore implement the postselection operator as $\Pi_f =
\int^{\Delta t} dt \, I(t)$.
In doing so we note that the detection of an electron in the drain at
time $t <\Delta t$ consists of a strong (postselection) measurement. 
Therefore, in assessing the weak value, we need to account only for
weak measurements that \emph{preceded} that strong measurement at time $t$.
This is implemented by constraining the correlation between $I(t)$ and
$N(s)$ for time intervals such that $s<t$.
Finally, since we are dealing with a steady state, the correlations
$\Avg{N(s) I(t)}$ depend only on the time difference $t-s$ and we can
write
\begin{equation}
\tau_{\textrm{WV}} = \frac{\int_0^{\infty} ds \,\Avg{ I(t) N(t-s) } }{ \Avg{I} } .
\end{equation}
The resulting  \emph{complex} $\tau_\textrm{WV}$ encodes the
information on the physical times involved in the cotunneling process.
Notably, the measured QPC-QD current correlation  provides access
only to a combination of the real and imaginary parts of
$\tau_\textrm{WV}$ (which depends on non-universal details of the
variation of the QPC’s transparency as function of the QD’s
occupation). The individual parts can be obtained through detailed
tomography of $\tau_\textrm{WV}$.
To shed light on the physical meaning of $\tau_\textrm{WV}$, it is
instructive to first discuss the analogous complex tunneling time in
the context of a single particle tunneling.

 \paragraph*{Analysis of a complex $\tau_\textrm{WV}$ for a single particle
   tunneling.}  Here we review the analysis of the tunneling time of a single particle
(\emph{single particle} opaque barrier), through a weak value
protocol. This serves as a benchmark for the analysis of the
equivalent cotunneling time.
The issue of the time of single particle tunneling has been discussed extensively in
the literature in a variety of approaches; both vis-a-vis single particle
tunneling in real
space~\cite{Condon1931,Wigner1955,Buttiker1982,Buttiker1983,Sokolovski1993}, 
and to Landau-Zener tunneling in energy space~\cite{Mullen1989}.
A weak measurement approach to this problem~\cite{Steinberg1995} gives rise
to a tunneling time and a reflection time, 
$\tilde{\tau}_{\textrm{tun}}$ and $\tilde{\tau}_{\textrm{ref}}$
respectively, both being complex.
The weighted dwell time under the barrier is then
$\tilde{\tau}_\textrm{dwell} =T
\tilde{\tau}_{\textrm{tun}} +  (1-T) \tilde{\tau}_{\textrm{ref}}$ ($T$
being the transmission probability). 
Quite remarkably, this last equality holds for the \emph{complex}
  tunneling and reflection times.
The physical times of this problem are  the dwell time, $\tilde{\tau}_{\textrm{dwell}
} = \mathrm{Re} \{ \tilde{\tau}_{\textrm{tun}} \}$, and the traversal
time under the barrier,
$\tilde{\tau}_{T} = [ \tilde{\tau}_{\textrm{dwell}}^2 +(\mathrm{Im} \{
\tilde{\tau}_{\textrm{tun}} \} )^2  ]^{1/2}$~\cite{Steinberg1995, Buttiker1983}.
In the limit of a high thin barrier $\tilde{\tau}_{\textrm{dwell}}$ vanishes
(most reflected particles spend a negligible time under the barrier),
hence $\tilde{\tau}_{T} = |\mathrm{Im} \{ \tilde{\tau}_{\textrm{tun}} \}|$. 
This agrees with the result of Ref.~\cite{Buttiker1983}.



\section{Microscopic calculation}
We now turn to calculate the cotunneling time in
\eq{eq:definizione}. 
\begin{figure}[ht!]
\begin{center}
\centerline{\includegraphics[width=80mm]{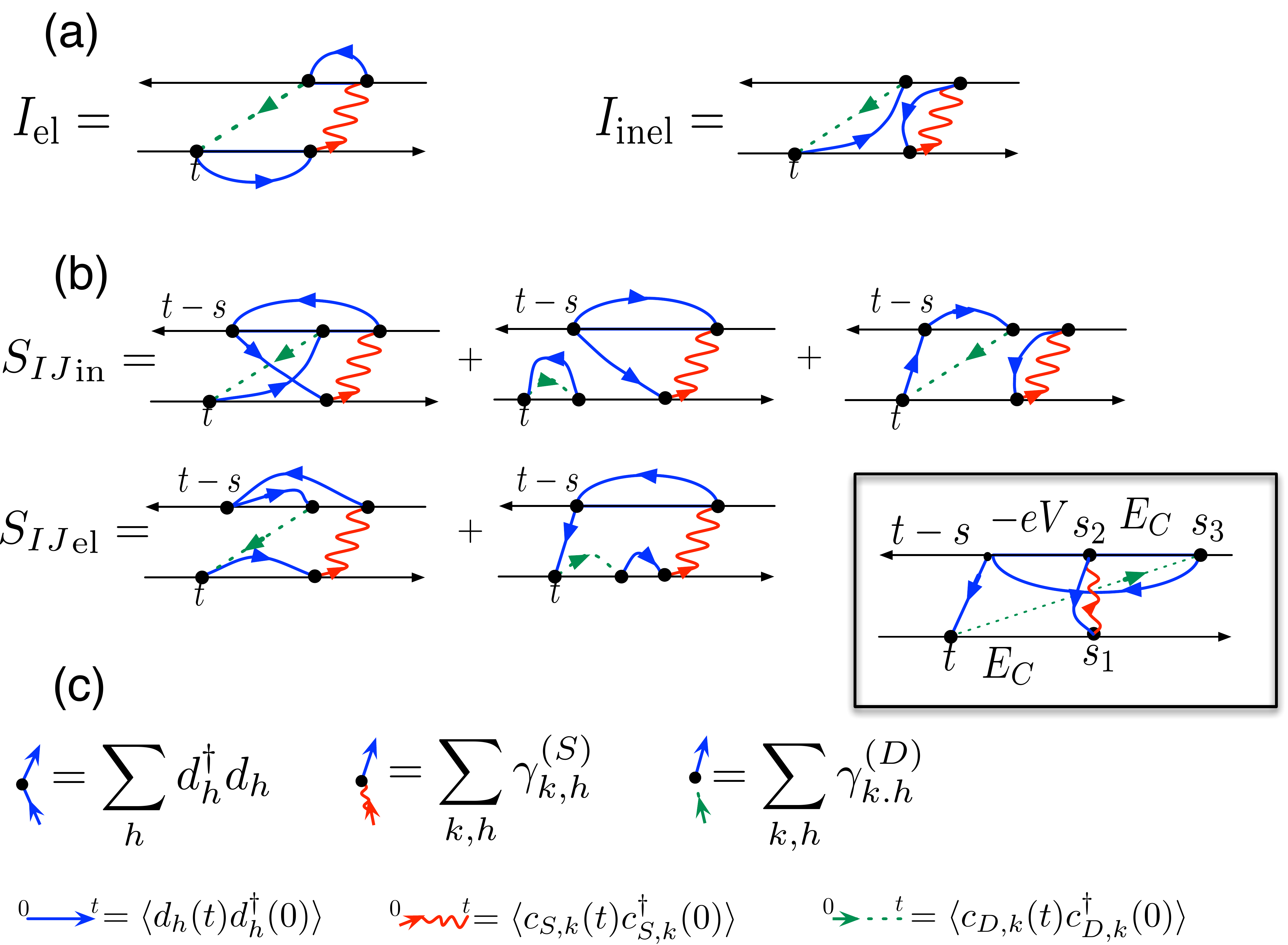}}
\caption{Feynman diagrams for $\Avg{I}$ and $S_{IJ}$. Elastic and
  inelastic contributions to the cotunneling current $\Avg{I}$
  [panel (a)], and
to the current-current correlator $S_{IJ}$ [panel (b)],  in the
  zero-temperature-particle-dominated regime. 
The propagators and vertices constituting the diagrams are defined in in panel (c).
 Inset:  An example of a diagram neglected in
  the zero-temperature-particle-dominated limit. Here we
  indicate explicitly the time and charging energy labelings along the
  time contour, as dictated  by the Feynman rules (cf.\ Appendix~\ref{feynmann}).}\label{fig:diagrammi}
\end{center}
\end{figure}

While the result for the cotunneling current is
known~\cite{Averin1990,Glazman2005}, it is  the charge-current correlation
function of the system that  encodes information on the
cotunneling time. 
For simplicity we tune the gate voltage such that transport through
the quantum dot is dominated by particle-like cotunneling. 
Our analysis addresses the limit where the temperature is
smaller than the source-drain voltage.
Different contributions to the current are described by Feynman
diagrams and are conveniently grouped into elastic and
inelastic contributions [cf.\ Fig.~\ref{fig:diagrammi} (a)]~\cite{Averin1990}.
Correspondingly, we group the  contributions to the correlator
$\Avg{I(t) [ N(t-s) -\Avg{N} ] }$ into two sets. 
The results for the complex cotunneling time will differ depending on
whether the cotunneling is dominated by elastic or inelastic
processes.
Examples of diagrams that contribute to $\Avg{I}$ and $\Avg{I(t)
  [ N(t-s) -\Avg{N} ] }$ are depicted in Fig.~\ref{fig:diagrammi}.
All in all we have 64 different diagrams; when dealing
with particle-dominated cotunneling and considering the
zero-temperature limit, the number of diagrams is reduced to 8  (4
elastic processes and 4 inelastic processes), as shown in
Appendix~\ref{the-qd}. The non-vanishing
diagrams in this limit are depicted in
Fig.~\ref{fig:diagrammi}(b), and are evaluated in
Appendix~\ref{calcolo finale}. The expressions for  $\Avg{I(t) [
  N(t-s) -\Avg{N} ] }$ and $\Avg{I}$ in
Eqs.~(\ref{eq:I-el1},\ref{eq:I-in1},\ref{eq:in-misura-in},\ref{elastico-misura1})
in Appendix B are
 finally substituted into \eq{eq:definizione}
to find 
\begin{align}
\tau_{\textrm{WV}} =\mp \frac{i}{2} \partial_{E_C} \ln I ,
\label{eq:log-der}
\end{align}
with $+$ and $-$ for inelastic and elastic cotunneling respectively.
An equivalent expression yielding  the traversal time in terms of the
logarithmic derivative of the transmission probability holds for the
non-interacting case, where, for tunneling through a square potential
barrier, it reads 
$\tau_{\textrm{WV}} = -\partial_{V_0} [\arg (t)]
+\frac{i}{2} \partial_{V_0} \ln (t^* t) $~\cite{Sokolovski1987}. Here $V_0$ is the barrier
height and $t$ the transmission amplitude.
Our result extends the validity of such an equation to interacting
systems.

Eq.~(\ref{eq:log-der}) is rather general, and does not depend on the
specifics of the electronic dynamics in the quantum
dot. However, in order to obtain specific expressions for $\tau_{\textrm{WV}}$,
we specify the dot to be in the diffusive limit, $L > |\mathbf{x_S} -
\mathbf{x_D}| \gg l$, where $L$ is the linear size of the quantum dot,
$l$ the elastic mean free
path, and $\mathbf{x}_S$, $\mathbf{x}_D$ the position of the source
and drain contacts respectively.
As shown in
Appendix~\ref{the-qd}, the cotunneling current can then be expressed in
terms of the diffuson propagator~\cite{Mirlin2000} between the source
and the drain points $\mathcal{D}_{\omega} (\mathbf{x}_S ,
\mathbf{x}_D) \equiv  \nu_0 \mathcal{S}^2 \Avg{\psi_\alpha(\mathbf{x}_S)   \psi^*_\alpha (\mathbf{x}_D)
 \psi_\beta(\mathbf{x}_D)   \psi^*_\beta (\mathbf{x}_S) }$, where
$\psi_\alpha(\mathbf{x})$ is the wave-function of the $\alpha$-energy
level of the dot with energy $\epsilon_\alpha$,
$\omega=\epsilon_\alpha-\epsilon_\beta$, $\nu_0$ and $\mathcal{S}$ are
respectively the density of
states and the area of the dot, and the average is intended
over the different statistical realizations of disorder.

In this case the
cotunneling current reads~\cite{Averin1990,Glazman2005,Mirlin2000,Aleiner1997}
\begin{align}
I_\textrm{in}  =&  \frac{G^{(S)} G^{(D)} }{12 \pi e^2} 
\frac{(eV)^2}{E_C^2} V ,
\label{in-final-1}\\
I_\textrm{el}  =&  \frac{G^{(S)} G^{(D)} }{4 \pi^2  \nu_0 e^2} V
 \int_{0}^{\infty} d \omega \,\frac{\mathcal{D}_\omega (\mathbf{x}_S ,
   \mathbf{x}_D) + \mathcal{D}_{-\omega} (\mathbf{x}_S ,
   \mathbf{x}_D)}{\omega} \nonumber \\
& \ln \left( 1+ \frac{\omega}{E_C} \right ) .
\label{el-final-1}
\end{align}
The elastic cotunneling current depends on the diffuson propagator,
which is characterized by Thouless energy, $E_\textrm{Th} \sim
D/\mathcal{S}$, proportional to the diffusion constant, $D$.
The cotunneling current depends on the ratio  between
$E_{Th}$ and $E_C$.
The cotunneling time in the various regimes is finally  
deduced directly  from \eq{eq:log-der} and the corresponding
expression for the cotunneling current,
Eqs.~(\ref{in-final},\ref{el-0d},\ref{el-long},\ref{el-short}). The
results are listed in Table~1.

We note here that, while the cotunneling current generally depends
parametrically on the various  energy scales in the different
regimes, the cotunneling time is $\tau_{\textrm{cot}} \sim \hbar/(E_C)$ in all
cases except for elastic cotunneling with very close contacts, $E_{\textrm{Th}} \ll E_C \ll \frac{L^2 E_\textrm{Th}}{
    |\mathbf{x}_S-\mathbf{x}_D|^2}$, where
it is given by  
\begin{align}
\tau_\textrm{WV} =- i\frac{3}{2 E_C} \, \ln^{-1}\left( \frac{D}{\pi^2
    |\mathbf{x}_s-\mathbf{x}_D|^2 E_C}\right) .
\end{align}
This shows explicitly that the cotunneling time can be parametrically
different from the estimation obtained via the uncertainty principle. 
In particular, the particle can spend a time shorter than $1/E_C$ in
the dot. This is in striking contrast with the results for tunneling
of a single particle through a barrier.

\begin{table*}[htb!]
\begin{tabular}{|l|c|c|c|}
\hline 
 & $\mathrm{Re} \{  \tau_{\textrm{WV}}\} = \tau_{\textrm{dwell}} $ &
 $\mathrm{Im} \{  \tau_{\textrm{WV}} \} = \pm \tau_{\textrm{cot}}$ 
 & $\Avg{I} / \left(  \frac{G^{(S)} G^{(D)} }{12 \pi e^2} V\right)$\\
\hline 
\hline
inelastic  &  0 & $ \frac{\hbar}{E_C} $  & $ \frac{ (eV)^2 }{ E_C^2} $   \\
\hline \hline
elastic & & & \\
\hline
$ E_C \ll E_{\textrm{Th}} $  & 0 & $ -\frac{1}{2}\frac{\hbar}{E_C}$
& $
\frac{3\delta }{E_c} $\\
\hline
{\footnotesize $E_{\textrm{Th}} \ll E_C \ll \frac{L^2 E_\textrm{Th}}{
    |\mathbf{x}_S-\mathbf{x}_D|^2}$, $2d$ }& 0  &   $ - \frac{3\hbar }{2 E_C} \,
\ln^{-1}\left( \frac{L^2 E_\textrm{Th}}{
    |\mathbf{x}_S-\mathbf{x}_D|^2 E_C}\right)$ & $ 
\frac{\delta }{E_\textrm{Th}} \ln^3\left( \frac{L^2 E_\textrm{Th}}{
    |\mathbf{x}_S-\mathbf{x}_D|^2 E_C} \right)$\\
\hline
 {\footnotesize$E_{\textrm{Th}} \lesssim\frac{L^2 E_\textrm{Th}}{
    |\mathbf{x}_S-\mathbf{x}_D|^2} \ll  E_C $, $2d$ } & 0  &   $ - \frac{ \hbar}{E_C}$ & $ 
\frac{3 \delta E_\textrm{Th}}{4 E_c^2} 
$\\
\hline
\end{tabular}
\caption{Real and imaginary components of the cotunneling time,
  $\tau_{\textrm{WV}}$ for the inelastic and elastic cotunneling
  regimes.
Listed are the respective average cotunneling currents too.
The results are written in terms of the dot-leads conductance
$G_\alpha = e^2 \nu_\alpha \nu_0 |\gamma^{\alpha}|^2/(2 \pi  \hbar)$,
with $\nu_{\alpha(0)}$ being the density of state in the lead (dot);
the relevant energy scales in the dot are the level spacing, $\delta$, the
Thouless energy, $E_{\textrm{Th}}$, the charging energy,  $E_{C}$, and
the applied voltage bias $eV$. $|\mathbf{x}_S -
  \mathbf{x}_D|$ is the distance between the source and drain
  contacts, $L$ the linear size of the quantum dot. The results in the last two rows are explicitly
    for the 2-dimensional case.}
\end{table*}

\section{Discussion and summary}
For a single particle tunneling through
an opaque barrier, it has been shown~\cite{Buttiker1983,Steinberg1995}
that the traversal time is given by the imaginary part of the complex
weak value time; The real part  turned out to be  the dwell time, and
was found to be vanishing. 
In the present analysis we  consistently find a vanishing dwell time
($\mathrm{Re}\{\tau_\textrm{WV}\}=0$), while $\tau_\textrm{cot}=
|\mathrm{Im} \{ \tau_\textrm{WV}\}|$ (cf.\ Table~1).
We also note that  Eq.~(\ref{eq:log-der}) has been established for
  single particle tunneling (opaque barrier~\cite{Sokolovski1987}). 
We have shown here that
  it remains valid for interacting systems.
The emerging picture shows that the cotunneling traversal time is not simply given
by the uncertainty time $\hbar/E_C$, but it can be
  logarithmically smaller in $E_\textrm{Th}/E_C$, $E_\textrm{Th}$ being
  the Thouless time-of-flight through the QD~\cite{Mirlin2000}
  (assuming, for example, diffusive dynamics in the dot).
The dependence on the ballistic or diffusive \emph{real}
time-of-flight through the QD is very different for the single
particle tunneling: the time of the latter is determined by the
imaginary velocity under the barrier~\cite{Buttiker1982}.

We have presented our analysis on three levels. 
First, we have defined a realistic system-detector setup both in the
sequential and cotunneling regimes, and related the
correlation function of the system-detector currents to
$\Avg{\tau_+^{(1)}}_\textrm{seq}$ or the complex
$\Avg{\tau^{(1)}_+}_\textrm{cot}$---Eqs.~(\ref{eq:tunneling-sequenziale},\ref{eq:tunneling-debole},\ref{eq:cotunneling}).
  These  expressions are useful for processing   experimental
  data,  through (i) analysis  of current-current  correlations
  [\eqs{eq:tunneling-debole}{eq:cotunneling}]  or 
  (ii), for single shot measurements, selective inclusion of  signals of  detector current
  conditional on the  later detection of a current pulse through the
  QD [\eq{eq:tunneling-sequenziale}].  
Second, through a weak value analysis, we have addressed the meaning
of the complex time $\tau_\textrm{WV}$.
This complete  $\tau_{\textrm{WV}}$  contains information about the
dwell and the cotunneling times.  
Third we have  considered
Eqs.~\ref{eq:tunneling-debole},~\ref{eq:cotunneling} as a starting
point for a first principles calculation of  $\tau_{\textrm{WV}}$,
pursued through a diagrammatic Keldysh  perturbation theory. 
Our protocol is amenable of experimental verification.
For a ballistic semiconducting QD whose  linear size is $L=0.15 \mu
m$,   the electron’s Fermi velocity $v_F=10^{6}cm/sec$  and  $E_C = 20
\mu eV$,   $E_{Th}$  and $E_{C}$ are comparable. One
may  design and tune the relevant gates to achieve the desired
inequality between these two energies.
Within a broader context, the analysis outlined here demonstrates the
usefulness of such composite measurements protocols for a systematic,
non-destructive  study of many-body systems driven to a virtual state.

\acknowledgments
We thank K. Ensslin, T. Ihn, and P. W. Brouwer for stimulating discussions.
We acknowledge the support of GIF, BSF under Grant No. 2006371, and
DFG under Grants No. RO 2247/8-1 and No. RO 4710/1-1.

\appendix

\begin{widetext}

\section{Sequential tunneling time: a classical case}
\label{sequenziale}

We discuss here the measurement of the sequential tunneling time 
through a weak measurement scheme.
 We show explicitly that the sequential tunneling time obtained
 through the weak  measurement scheme as in Eq.~(\ref{eq:tunneling-debole}) 
 coincides with the result of a direct strong measurement, Eq.~(\ref{eq:tunneling-sequenziale}).

To this goal we consider a simple model of transport through the
QD, where tunneling of subsequent electrons are uncorrelated
events.  
We assume a constant flux, $f_0$, of electrons emitted from the source. For
simplicity we assume that one electron impinges on the dot in the
time $T$, i.e. $f_0=1/T$; hereafter  $\Gamma_{S}$,
$\Gamma_{D}$ are the dot-source and dot-drain tunnel rates
respectively. 

To begin, we evaluate the probability of the events describing
transport in the dot.
Assuming that an electron enters the dot at $t=0$, and that the time for
tunneling out is Poissonian distributed, the probability, $P(t)$ that
an electron remains in the dot until time $t$ is 
\begin{align}
P(t) = e^{-(\Gamma_S +\Gamma_D) t} .
\end{align}
It follows that the probability, $p(t) \, dt$, that an electron remains in the dot till time $t$ and
exits it in the time interval $[t,t+dt]$ is given by
\begin{equation}
	p(t) \, dt =e^{-(\Gamma_S +\Gamma_D) t} (\Gamma_D+\Gamma_S) \,
        dt .
\end{equation}
We notice also that $P(D|t) = \Gamma_D /(\Gamma_S+\Gamma_D)$, where
$P(D|t)$ is the  probability that, given that the particle exits the
dot within the time interval $[t,t+dt]$, it does it through the drain.
The analogous equation with $\Gamma_D \leftrightarrow \Gamma_S$  holds
for the corresponding probability, $p(S|t) \, dt$, in  the case of
electron exiting toward the source.  
The probability that the electron exits the dot in the time interval $[t,t+dt]$,
given that this takes place through the drain's barrier, is determined
through the Bayes theorem, leading to $p_D(t) \, dt \equiv p(t|D) \, dt =
p(D|t) p(t)/P(D)$, where $P(D)=\Gamma_D/(\Gamma_S+\Gamma_D)$ is the
total  probability to exit to the drain, integrated over all times. 
This results in
\begin{equation}
	p_D(t) dt=  p(t) dt = ^{-(\Gamma_S +\Gamma_D) t} (\Gamma_S+\Gamma_D) dt .
\end{equation}
Note that an identical expression is obtained for $p_S(t)$.
The time the particle spends in the dot, given that it eventually
tunnels to the drain, is then 
\begin{align}
\Avg{\tau_+^{(1)}} \equiv \int_0^{\infty} dt \, t \, p(t|D) =
\frac{1}{\Gamma_D+\Gamma_S} \, .
\label{tempo-1}
\end{align}
We conclude that the time the electron spends in the dot is
independent of the condition of eventually exiting to the left or to the right.

Assessing the sequential tunneling time, \eq{tempo-1},  by employing
strong measurement protocol  is straightforward.
 One can correlate the entry of an electron to the QD (detected
 through a clear signal in the QPC~\cite{Sukhorukov2007}) with the
 detection of the electron at the drain ---the relevant quantity is
 $p(t|D)$.  
This becomes trickier when weak measurement by the detector is
employed. 
One then needs to resort to Eq.(~\ref{eq:tunneling-debole}), invoking the
correlation function $S_{IJ}$.  
We can determine the time resulting from Eq.~(\ref{eq:tunneling-debole})
assuming, without loss of generality,  that the current $J^{(1)}(s)=J + \xi(s)$ 
when the electron is in the dot, and $J^{(0)} (s)=0+\xi(s)$ otherwise
(a constant reference current has been subtracted).
$\xi(s)$ is a stochastic component of the current due to the
detector's intrinsic noise with $\int_{-\infty}^t ds \, \xi(s) =0$, and it
is uncorrelated to the QD signal.
We also note that the event of having a current peak in the drain at time $t$
happens with probability $P(D|t)$ and the average in $S_{IJ}$ has to be
taken with respect to this probability.
Measuring the current through the QD in units of the electron charge, we can write Eq.~(\ref{eq:tunneling-debole}) as
\begin{align}
S_{IJ} = P(t=0) \int_0^\infty dt \, \int_0^t ds\,  [J P(D|t) p(t) +
\xi(s)]  = (\Gamma_S /f_0) J \int_0^\infty dt\, t\,
\Gamma_D \, e^{(\Gamma_D+\Gamma_S) t } \, ,
\label{tempo-2}
\end{align}
where $P(t=0)=\Gamma_S /f_0$ is the probability of entering the dot
at time $t=0$.
In the first equality of \eq{tempo-2} the term involving $\xi(s)$ is
not weighted by any probability since it is not correlated to the
dynamics of the dot.
In the second equality in \eq{tempo-2} we take into account that
the time integral of that same term involving the 
stochastic fluctuations vanishes.
We may further write $I = \Gamma_S \Gamma_D/(\Gamma_S+\Gamma_D)$.
With this expression, Eq.~(\ref{eq:tunneling-sequenziale}) leads to 
\begin{align}
\Avg{\tau_+^{(1)}} = 1/(\Gamma_S+\Gamma_D),
\end{align}
in full agreement with the result of a strong measurement.

\section{Calculation of the QD correlation function}
\label{the-qd}

In this section we discuss the calculation of $\Avg {I}$ and
$\Avg{I(t) [N(t-s)-\Avg{N} ] }$ in the electron cotunneling regime. 
We address first the calculation of the average current, and then its
generalization to the charge-current correlation function.

\subsection{Average current}
The Hamiltonian is presented in Section~\ref{modello}.
The calculations are done perturbatively in the tunneling Hamiltonian,
$H_T$, hence we work in the interaction picture with respect to
$H_0+H_\textrm{int}+H_\textrm{QPC}$, and denote by $\bar{\cdot}$ the operators in the
interaction picture. 
The leading order term in the perturbative calculation of the current
is obtained to third order in $H_T$. 
The cotunneling current~\cite{Averin1990}  in \eq{corrente}  is in turn proportional to
$\gamma^4$:
\begin{align}
\Avg{I} = & e \, \mathrm{Re} \left\langle  
 \int_{-\infty}^{t} ds_1 \,
    \int_{-\infty}^{s_1} ds_2 \,  \int_{-\infty}^{s_2} ds_3 \, 
\left[ 
\bar{T}_D(t) \bar{H}_T(s_1) \bar{H}_T(s_2) \bar{H}_T(s_3) +
 \bar{H}_T(s_3) \bar{H}_T(s_2) \bar{H}_T(s_1) \bar{T}_D(t) 
\right]  \right. \nonumber\\
&  - \left.  \int_{-\infty}^{t} ds_1 \, \int_{-\infty}^{t} ds_2 \,  \int_{-\infty}^{s_2} ds_3 \, \left[ 
 \bar{H}_T(s_1) \bar{T}_D(t)  \bar{H}_T(s_2) \bar{H}_T(s_3) +
 \bar{H}_T(s_3) \bar{H}_T(s_2) \bar{T}_D(t) \bar{H}_T(s_1) 
\right]
\right\rangle,
\label{eq:perturbazione}
\end{align}
where the average is intended on an unperturbed state, (possibly a
mixed one), described by a density matrix $\rho (t= - \infty)$.

Each of the $H_T$ terms in \eq{eq:perturbazione} consists in fact of a sum of
terms involving products of $T_S$, $T_D$.
The system's state is initially (at $t=-\infty$) an eigenstate
of $N$, $N_S$, $N_D$. It remains an eigenstate of the same operators
after the application of each $H_T$ (note that $[N,T_S]=T_S$,
$[N,T_S^\dag]=-T_s^\dag$, $[N,T_D]=T_D$, $[N,T_D^\dag]=-T_D^\dag$).  
In fact $H_T$ changes the system's  state to a configuration with $\pm 1$
extra charge on the QD.
The time dependence in the operators in \eq{eq:perturbazione} can then be
made explicit in terms of the time evolution operator and computed
employing \eq{eq:energia-carica}. 
To be specific, let us address one of the terms appearing in \eq{eq:perturbazione}, namely
\begin{align}
\label{eq:esempio}
\Avg{I}_1\equiv e \, \mathrm{Re} \langle  
 \int_{-\infty}^{t} ds_1 \,    \int_{-\infty}^{s_1} ds_2 \,  \int_{-\infty}^{s_2} ds_3 \, 
\bar{T}_D(t) \bar{T}_S^\dag(s_1) \bar{T}_D^\dag(s_2)
\bar{T}_S(s_3)\rangle \,.
\end{align}
The explicit time dependence of operators results then in
\begin{align}
\Avg{I}_1= e \,\mathrm{Re} & \mathrm{Tr} \left\{ 
 \int_{-\infty}^{t} ds_1 \,    \int_{-\infty}^{s_1} ds_2 \,  \int_{-\infty}^{s_2} ds_3 \, 
 T_D (t)\,  e^{-i (E_{+l}+eV) (t-s_1)}
T_S^\dag (s_1) \, \right. \nonumber\\
 & \left. e^{+ i eV (s_1-s_2)} \, 
T_D^\dag (s_2) \, e^{-i E_{+l} (s_2-s_3)} \, 
T_S (s_3) \rho(-\infty)\right\} \, ,
\label{eq:esempio1}
\end{align}
where the operators evolve in time through 
\begin{align}
\mathcal{U}(t,t') =
\mathcal{T}-\exp\{-i H_0   (t-t')\},
\label{yu3}
\end{align} 
 where $\mathcal{T}$ is  the time
ordering operator, and $H_0$ is the Hamiltonian  of the
dot-leads part, which includes neither the charging energy, nor
the detector Hamiltonian.
The fact that the detector Hamiltonian is neglected in \eq{yu3} is a
consequence of the perturbative calculation in the strength of the
measurement. More precisely, since Eq.~(\ref{eq:misura-QPC}) is
obtained by already computing $S_{IJ}$ to first order in $H_{\textrm{int}}$,
the $S_{IJ}$ can be safely neglected in the calculation of $\Avg{I(t)
  [ N(t-s)- \Avg{N} ] }$. 
A proper treatment of $H_{\textrm{int}}$ in the calculation of
$\Avg{I}$ reveals that it makes contributions in second order, and is therefore consistently neglected here~\cite{Romito2014}. 
 
The quantum
average in \eq{eq:esempio} can now be easily obtained via  Wick's theorem.
This is conveniently done in terms of Feynman diagrams on  Keldysh
contour, resulting in the rules specified in Appendix~\ref{feynmann}. 
All possible diagrams correspond to all possible sequences of
$T_{\alpha}(s_j)$ obtained from the $H_T$ in \eq{eq:perturbazione},
and are presented in~\fig{all-I}.

\begin{figure*}
\includegraphics[width=140mm]{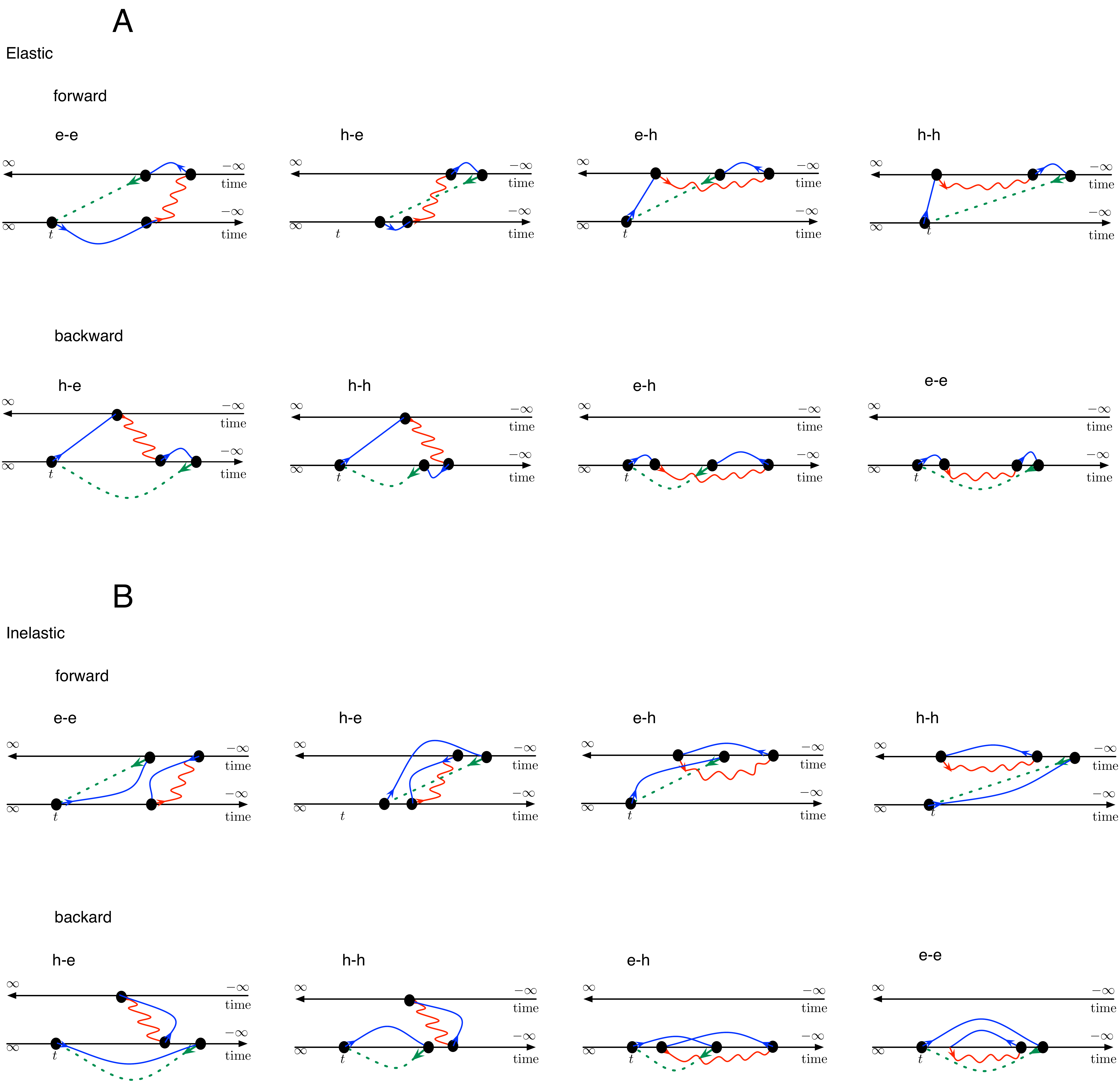}
\caption{Feynman diagrams contributing to the cotunneling
  current. Each contribution to the probability consists of a coherent
  superposition of electron-like (e) and hole-like (h)  amplitudes. Upon
  squaring these amplitudes one obtains contributions of the type e-e,
  e-h, etc., which are represented by the various diagrams.  
The diagrams are grouped into elastic (A) and inelastic (B)
  contributions.
Each group consists of ``'forward'' and ``backward''
diagrams, depending on whether the associated charge transfer is from source to drain or viceversa, respectively.
The contribution of the latter set of
diagrams is vanishing at $T=0$.
The relative weight of the electron and hole contributions is
controlled by the gate voltage.
}
\label{all-I}
\end{figure*}

Each diagram corresponds to a well defined process which contributes to the
\emph{probability} of charge transfer  between the two leads. 
The \emph{amplitude} for a charge transfer is a superposition of electron-like (e)
and hole-like (h) processes. 
The various contributions to the probability are labeled accordingly
(e.g., $e-e$, $e-h$, $\dots$).
Moreover the diagrams are classified as forward (backward)
when a charge is transferred from the source (drain) to the drain
(source).
For instance, the diagrams obtained by averaging the expression for
$\Avg{I}_1$ in \eq{eq:esempio} are the ``e-h forward'' both elastic and inelastic,
according to the labeling in~\fig{all-I}
.

The analytical expression for each diagram is obtained by the rules
R7-R10 in Appendix~\ref{feynmann}. 
The backward diagrams are vanishing in the $T=0$ limit due to the
vanishing of the
corresponding phase space. 
Moreover, by controlling the gate voltage, one can tune the ratio
$E_{-r}/E_{+l}$ in the regime $E_{-r}/E_{+l} \ll 1$, where the
hole-like processes are parametrically suppressed by a factor
$\mathcal{O}(E_{-r}/E_{+l})$.
In this zero-temperature particle-dominated limit, the elastic and
inelastic contributions to the current are given by
\begin{align}
I_\textrm{el} =\includegraphics[width=20mm]{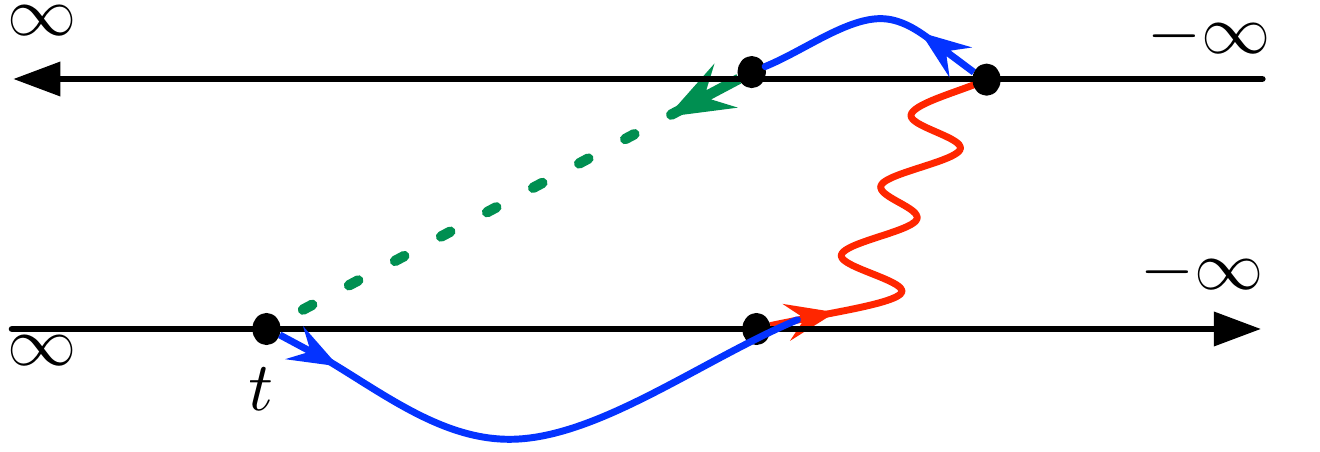}  +
\textrm{h.c.} = & - 2 e \,
 \mathrm{Re} \left\{ \int_{\mathbb{R}_+^3 }  dx dy dz \,
\sum_{\alpha,\beta, j, k} 
(1-n_\alpha) (1-n_j) (1-n_\beta) \, n_k \, 
\gamma^{(S)}_{\alpha, k} { \gamma^{(D)}_{\alpha, j}}^*
\gamma^{(D)}_{\beta, j} { \gamma^{(S)}_{\beta, k}}^* \right. \nonumber \\
 &   \left. \phantom{\frac{a^a}{a^a}}e^{-i (E_{+,l} +\epsilon_\alpha -\epsilon_k ) x } e^{-i (- eV +
  \epsilon_j -\epsilon_k ) y } e^{i ( E_{+,l} +\epsilon_\beta -\epsilon_k ) z} 
\right\} ,  \label{eq:I-el}\\
I_\textrm{in} =  \includegraphics[width=20mm]{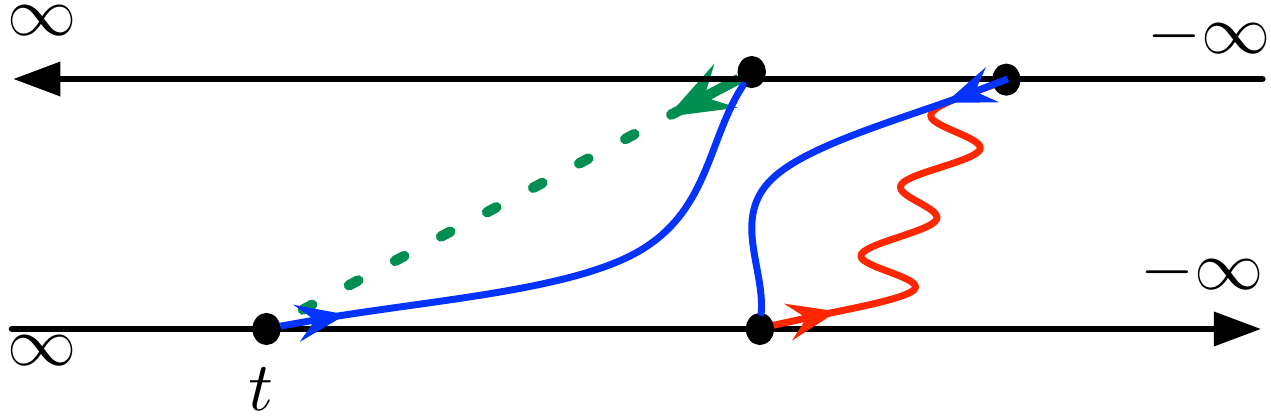}  +
\textrm{h.c.} = &  - 2 e \,
 \mathrm{Re} \left\{ \int_{\mathbb{R}_+^3 }  dx dy dz \,
\sum_{\alpha,\beta, j, k} 
(1-n_\alpha) (1-n_j) n_\beta \, n_k \, 
| \gamma^{(S)}_{\alpha, k} |^2  | \gamma^{(D)}_{\beta, j} |^2 \right. \nonumber \\
 &   \left. \phantom{\frac{a^a}{a^a}}e^{-i (E_{+l} +\epsilon_\alpha
     -\epsilon_k ) x } e^{-i (- eV + \epsilon_\alpha  -\epsilon_\beta
     + \epsilon_j -\epsilon_k ) y } e^{i ( E_{+l} +\epsilon_\alpha
     -\epsilon_k ) z}  
\right\} , \label{eq:I-in}
\end{align} 
with $n_\eta$ the distribution function of the occupation of the
$\eta$-th energy level. 

To proceed further one notes that the tunneling matrix elements can be
written in terms of the dot's and leads' wave functions, $\psi(\mathbf{x})$
and $\phi(\mathbf{x})$ respectively, as  $\gamma_{\alpha,k}^{(S)} =
\sqrt{\mathcal{V}\mathcal{S}} \gamma
\psi_\alpha(\mathbf{x}_S) \, \phi_k(\mathbf{x}_S)$, where
$\mathbf{x}_S$ is the coordinate of the tunneling point between the
source and the dot,  $\mathcal{V}$ the volume of each lead, and
$\mathcal{S}$ that of the dot.  
An  equivalent expression holds for the drain.
These are realization dependent quantities, and we consider their
statistical average.
Independently of the QD dynamics, $ \mathcal{V} \Avg{  {\phi_k^{(S)}}^*
  (\mathbf{x}_S) \phi_k^{(S)} (\mathbf{x}_S) } = 1$.
Introducing the density of states in the leads, $\sum_k \to
\int_\mathbb{R} d\epsilon_k \, \nu$, and in the dot, $\sum_\alpha \to \int_{\mathbb{R}}
d\epsilon_\alpha \, \nu_0$, the integrals in Eqs.~(\ref{eq:I-el},\ref{eq:I-in}) are
finally performed to obtain
\begin{align}
I_\textrm{in} = & e \frac{G^{(S)} G^{(D)}}{2 \pi  e^4} \,
\int_{\mathbb{R}+^4}  d \epsilon_\alpha d \epsilon_\beta d
\epsilon_j d \epsilon_k \frac{\delta (\epsilon_\alpha + \epsilon_\beta +
\epsilon_j + \epsilon_k -eV ) \mathcal{S}^2 \Avg{\psi_\alpha(\mathbf{x}_S)   \psi^*_\alpha (\mathbf{x}_S)  \psi_\beta(\mathbf{x}_D)   \psi^*_\beta (\mathbf{x}_D) }}{ (E_C +\epsilon_\alpha +\epsilon_k )^2
}, \label{eq:I-in1}\\
I_\textrm{el} = & e  \frac{G^{(S)} G^{(D)}}{2 \pi  e^4} eV \,
\int_0^\infty d\epsilon_\alpha \, \int_0^\infty d\epsilon_\beta
\frac{\mathcal{S}^2 \Avg{\psi_\alpha(\mathbf{x_S})   \psi^*_\alpha (\mathbf{x}_D)
  \psi_\beta(\mathbf{x_D})   \psi^*_\beta (\mathbf{x}_S) }}{(E_C+\epsilon_\alpha)(E_C+\epsilon_\beta)}
+\mathcal{O}((eV)^2), \label{eq:I-el1}
\end{align}
where $G^{(a)} \equiv e^2 \nu \nu_0 |\gamma^{(a)}|^2/ (2 \pi \hbar)$
is the conductance of the $a=S,D$ contact

The averages over the statistical realizations
in Eqs.~(\ref{eq:I-el1},\ref{eq:I-in1}) are well-known in the
literature~\cite{Mirlin2000}. 
In the diffusive limit  $L > |\mathbf{x_S} - \mathbf{x_D}| \gg l$,  $
\mathcal{S}^2 \Avg{ {\psi_\alpha}^* (\mathbf{x}_S) 
\psi_\alpha (\mathbf{x}_S) {\psi_\beta}^* (\mathbf{x}_D)
\psi_\beta (\mathbf{x}_D)} \approx 1$
 , and 
 $\mathcal{S}^2 \Avg{\psi_\alpha(\mathbf{x}_S)   \psi^*_\alpha (\mathbf{x}_D)
 \psi_\beta(\mathbf{x}_D)   \psi^*_\beta (\mathbf{x}_S) } \equiv
\mathcal{D}_{\omega} (\mathbf{x}_S , \mathbf{x}_D)/\nu_0$,
 where
$\mathcal{D}_{\omega} (\mathbf{x}_S , \mathbf{x}_D)$ is the diffuson
propagator~\cite{Mirlin2000} between the source and the drain points,
and $\omega=\epsilon_\alpha-\epsilon_\beta$.
The cotunneling current then reads~\cite{Averin1990,Glazman2005,Mirlin2000,Aleiner1997}
\begin{align}
I_\textrm{in}  =&  \frac{G^{(S)} G^{(D)} }{12 \pi e^2} 
\frac{(eV)^2}{E_C^2} V ,
\label{in-final}\\
I_\textrm{el}  =&  \frac{G^{(S)} G^{(D)} }{4 \pi^2  \nu_0 e^2} V
 \int_{0}^{\infty} d \omega \,\frac{\mathcal{D}_\omega (\mathbf{x}_S ,
   \mathbf{x}_D) + \mathcal{D}_{-\omega} (\mathbf{x}_S ,
   \mathbf{x}_D)}{\omega} \ln \left( 1+ \frac{\omega}{E_C} \right ) .
\label{el-final}
\end{align}

The elastic cotunneling current depends on the diffuson propagator,
which is characterized by Thouless energy, $E_\textrm{Th} \sim
D/\mathcal{S}$, proportional to the diffusion constant, $D$.
The cotunneling current depends on the ratio  between
$E_{Th}$ and $E_C$.
In the limit $E_\textrm{Th} \gg E_C$ the elastic cotunneling current
acquires the universal form
\begin{align}
I_\textrm{el,0D}  =  \frac{G^{(S)} G^{(D)} }{4 \pi e^2}
\frac{\delta}{E_C} V . \label{el-0d} 
\end{align}
The expression is independent on the dot parameters and dynamics, and
can be regarded as the expression for a ``zero-dimensional'' dot.
In the opposite limit, $E_\textrm{Th} \ll E_C$, the result depends on
the dot's shape and the electron's dynamics therein.

Addressing now  the cotunneling time in the latter regime
we consider the cotunneling current in the specific case of a square
dot of linear size $L$. 
We expect our result to be parametrically correct for other dot's shapes.
The Thouless energy is then $E_{\textrm{Th}} = D/(\pi^2 L^2)$ and the
diffuson is expressed by
\begin{align}
\mathcal{D}_\omega (\mathbf{x}_S ,
   \mathbf{x}_D) =\sum_{\mathbf{n} \in {\mathbb{N}^+}^d } \frac{S^{-1} \Phi(\mathbf{x}_S ,
   \mathbf{x}_D)}{-i \omega +
     E_{\textrm{Th}} |\mathbf{n}|^2},
\end{align}
where $\mathbf{n}^T=(n_1, ..., n_d)$ in $d$ spatial dimensions, and
$\Phi(\mathbf{x}_S ,   \mathbf{x}_D) = (1/L^d) \prod_{j=1}^d
\cos(\frac{n_j x_{S,j}}{\pi L}) \cos(\frac{n_j x_{D,j}}{\pi L})$.
We focus here on the case of a two dimensional dot, though the
calculation can be performed in general in any spatial
dimensions~\cite{Romito2014}; we assume for simplicity
$\mathbf{x}_s=\mathbf{0}$.
 The parameter $\eta \equiv \pi |\mathbf{x}_S-
 \mathbf{x}_d|\sqrt{E_C/D}$ discriminates between the two regimes
 $l/L \ll \eta \ll1$ and $l/L \ll 1 \ll \eta$.
In the latter case the cotunneling current is known to be~\cite{Averin1990}
\begin{align}
I_\textrm{el,long}  =&  \frac{G^{(S)} G^{(D)} }{16 e^2} V \frac{\delta E_\textrm{Th}}{E_C^2}.\label{el-long}
\end{align}
This limit corresponds to the case of the source and drain contact at
the opposite sides of the dot.
In the opposite regime of source and drain close to each other we
estimate the cotunneling current as
\begin{align}
I_\textrm{el,short}   \approx &  \frac{G^{(S)} G^{(D)} }{8 \pi e^2} V \frac{\delta
}{E_\textrm{Th}} \int_0^\infty \frac{dx}{x} J_0(\eta x) f(x) .
\end{align}
Here $J_0 (x)$ is the $0$-th Bessel function, 
\begin{align}
f(x) = Li_2\left(\frac{1}{1-i x}\right) +Li_2 \left(\frac{1}{1-i
    x}\right) + 2 \pi \arctan(x)-\arctan^2(x)-\frac{\pi^2}{3},
\end{align}
and $-Li_2(-x) = \int_0^x dy \, \ln(1+y)/y $ is the dilogarithmic function.
We are interested in the limit $\eta \ll 1$ where 
\begin{align}
I_\textrm{el,short}  \approx &  \frac{G^{(S)} G^{(D)} }{12 \pi e^2} V
\frac{\delta }{E_\textrm{Th}} \ln^3\left( \frac{D}{\pi^2 |\mathbf{x}_s-
  \mathbf{x}_D|^2 E_C }\right). \label{el-short}
\end{align}

\subsection{Charge-current correlation function $\Avg{I(t) N(t-s)}$}
\label{calcolo finale}

The calculation presented above for the cotunneling current can be
easily generalized to the correlation function $\int_0^\infty ds \, \Avg{I(t) ( N(t-s) -\Avg{N})}$. 
Given the specific time order between the operators, we focus on
the calculation of $ \mathcal{F} (s)
=\Avg{\mathcal{T}_K [I_-(t) (N_+(t-s) -\Avg{N} ) ] } $ written
in terms of Keldysh time-ordered operators, and address the
time integral later.

The calculation is done perturbatively in $\gamma$, in complete analogy
with  the case of the current.
This leads to Feynman diagrams constructed according to the same rules
R1-R6 discussed above. 
In fact, the diagrams obtained for the correlator can be easily deduced
from the diagrams of the current. 
Each diagram contributing to  the cotunneling current has its analog for
the correlator at hand; the only difference between the two is the
insertion of the vertex $N_+(t-s)$ in the upper branch. 
As presented in~\fig{fig:I-to-IN},
this has two consequences: 
(i) the new $N$ vertex added to a certain current diagram can be
connected in two  different ways, leading correspondingly to two distinct
 diagrams for the correlation function;   
(ii) two diagrams of $I$ related by complex conjugation, become no longer
the complex conjugate of each other when inserting the new $N$
vertex, so they have to be computed separately.

\begin{figure*}
\includegraphics[width=140mm]{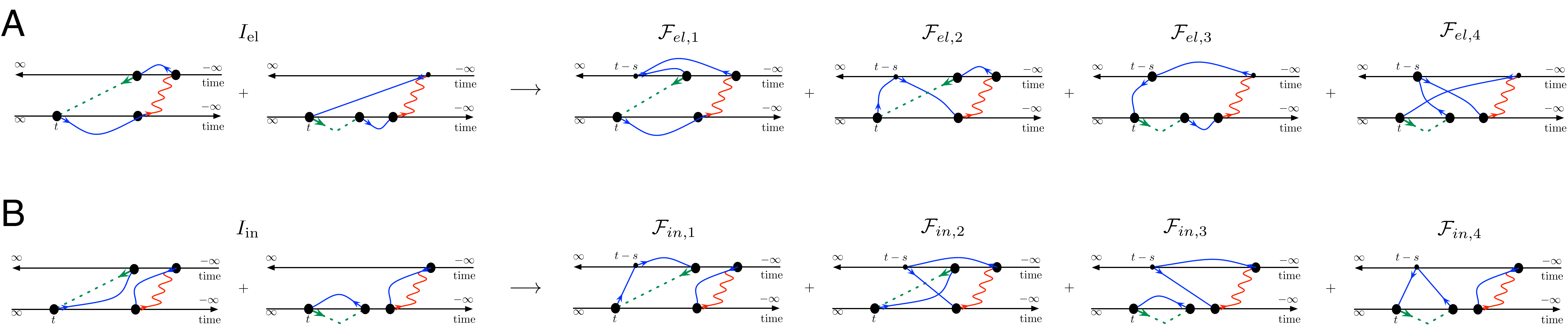}
\caption{Example of Feynman diagrams for the correlation function $\Avg{I_+(t)
     [ N_-(t-s) -\Avg{N} ] }$ obtained from the corresponding diagrams for the
  cotunneling current. The presented diagrams are all those contributing the
  correlation function in the limit of zero-temperature and
  particle-dominated processes in the elastic (A) and inelastic (B) cotunneling
  regimes. For each diagram contributing the cotunneling
  current there are four diagrams contributing the correlation
  function.} 
\label{fig:I-to-IN}
\end{figure*}

All the non-vanishing diagrams in the limit of zero-temperature and
particle dominated cotunneling are depicted in~\fig{fig:I-to-IN}
. 
Technically, the  new propagator including an $N$
vertex, namely $\Avg{\mathcal{T}_K [c_{\alpha \pm}(s_1) (N_+(t-s) -\Avg{N})
  c_{\beta,\pm}^\dag(s_2)]}$, can be directly evaluated. In the Schr\"odinger picture
\begin{align}
& \Avg{ c_{\alpha} (N -\Avg{N})
  c_{\beta}^\dag}  = \delta_{\alpha,\beta} (1-n_\alpha)^2=
\delta_{\alpha,\beta}(1-n_\alpha) , \label{eq:propag1}\\
&\Avg{ c_{\alpha}^\dag (N -\Avg{N})
  c_{\beta}}  = - \delta_{\alpha,\beta} n_\alpha^2= -
\delta_{\alpha,\beta} n_\alpha^2 , \label{eq:propag2} \\
&\Avg{ c_{\alpha}^\dag (N -\Avg{N})
  c_{\beta}^\dag}  = \Avg{ c_{\alpha} (N -\Avg{N})
  c_{\beta}}=0 ,
\end{align}
where the last equalities in Eqs.~(\ref{eq:propag1},\ref{eq:propag2}) are valid at
$T=0$.
In fact, the above equations can be effectively implemented in the
Feynman diagrams as expressed in the rule R10.
It immediately follows that, among all the contributing diagrams in~\fig{fig:I-to-IN}
,
$\mathcal{F}_{\textrm{el},2}
=\mathcal{F}_{\textrm{el},4}=\mathcal{F}_{\textrm{in},4}=0$. 
As an example, the non-vanishing diagram $\mathcal{F}_{\textrm{in},1} (s)$
in~\fig{fig:I-to-IN}
 reads:
\begin{align}
\includegraphics[width=20mm]{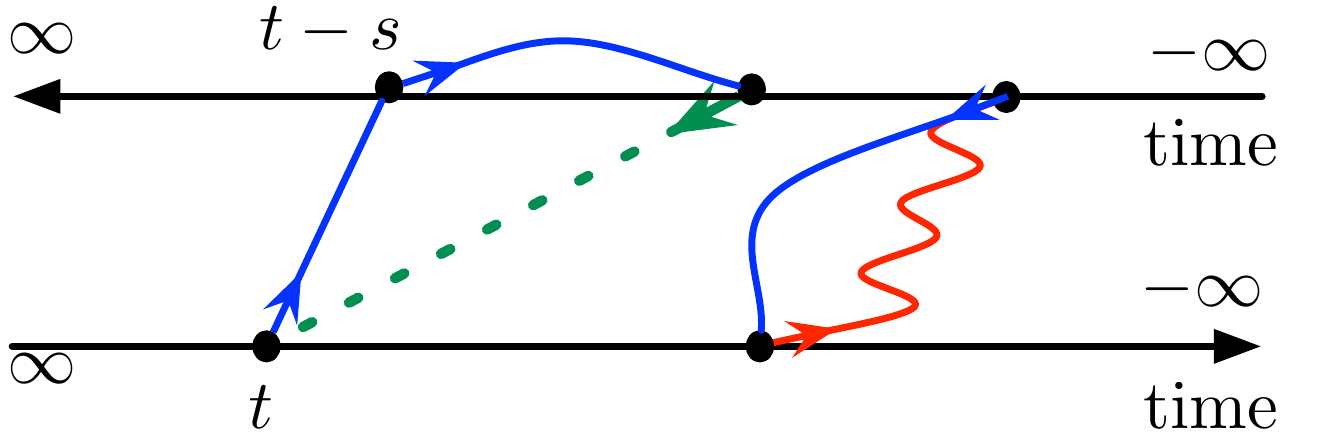}  & =  e \, 
\int_{0}^\infty  dx \, \int_0^\infty dy \, \int_0^{\infty }dz \,
\sum_{\alpha,\beta, j, k} 
(1-n_\alpha) (1-n_j) \, n_\beta^2 \, n_k \, 
|\gamma^{(S)}_{\alpha, k}|^2 |{\gamma^{(D)}_{\alpha, j}}|^2 \nonumber \\
 &   e^{-i (E_{+,l} +\epsilon_\alpha +\epsilon_k ) x } 
e^{-i (- eV +  \epsilon_\alpha + \epsilon_k  +\epsilon_\beta+\epsilon_j
  ) y } 
e^{i ( E_{+,l} +\epsilon_\alpha +\epsilon_k ) z}  
e^{-i (- eV +  \epsilon_\alpha + \epsilon_k  +\epsilon_\beta+\epsilon_j
  ) s }.
\end{align}

A direct evaluation of all the diagrams in the inelastic and elastic regime, done
in complete analogy with the calculation of the current, leads to
\begin{align}
\mathcal{F}_\textrm{in} (s) =  & \mathcal{F}_{\textrm{in},1} (s)+
\mathcal{F}_{\textrm{in},2} (s) + \mathcal{F}_{\textrm{in},3} (s)= 
 -i e\,\, \frac{G^{(S)} G^{(D)}}{2 \pi e^4} 
\int_{\mathbb{R}_+^4} d \epsilon_\alpha d \epsilon_\beta d \epsilon_j
d \epsilon_k \, 
L^2   \Avg{\psi_\alpha(\mathbf{x}_S)   \psi^*_\alpha (\mathbf{x}_S) 
  \psi_\beta(\mathbf{x}_D)   \psi^*_\beta (\mathbf{x}_D) }  
 \nonumber\\
& \frac{e^{-i ( E_C+\epsilon_\alpha+\epsilon_k)s } }{ ( E_C +
  \epsilon_\alpha +\epsilon_k)^2 } \left[
  \frac{1}{\epsilon_\alpha + \epsilon_\beta +\epsilon_j +
    \epsilon_k -eV +i\zeta } - \frac{1-e^{-i ( \epsilon_\beta
      +\epsilon_j -E_C -eV) s} }{\epsilon_\beta     +\epsilon_j -E_C
    -eV} \right] , \label{eq:F-in}\\
\mathcal{F}_\textrm{el} (s) =  &\mathcal{F}_{\textrm{el},1} (s)+
\mathcal{F}_{\textrm{el},3} (s) =  i e\, 
\frac{G^{(S)} G^{(D)}}{2 \pi e^4}\, \int_{\mathbb{R}_+^4} d
\epsilon_\alpha d \epsilon_\beta d \epsilon_j d \epsilon_k \, L^2
  \Avg{\psi_\alpha(\mathbf{x}_S)   \psi^*_\alpha (\mathbf{x}_D) 
  \psi_\beta(\mathbf{x}_D)   \psi^*_\beta (\mathbf{x}_S) }   \nonumber\\ 
 & \frac{e^{-i ( E_C+\epsilon_\alpha + \epsilon_k ) s } }{ ( E_C +
   \epsilon_\beta +\epsilon_k) (E_C + \epsilon_\alpha +\epsilon_k) }  
\left[  \frac{1}{\epsilon_j +
    \epsilon_k -eV +i\zeta } + \frac{1-e^{-i ( \epsilon_\alpha
      +\epsilon_j -E_C -eV) s} }{\epsilon_j - \epsilon_\alpha -E_C
    -eV}\right]. \label{eq:F-el}
\end{align}
with an infinitesimal regularization parameter $\zeta$.
After integrating over $s$ (eventually including a convergence factor
$e^{-\zeta s}$), we obtain
\begin{align}
\int_0^\infty ds \, \mathcal{F}_\textrm{in} (s) = &   i e \, \frac{G^{(S)} G^{(D)}}{2 \pi
  e^4}  \, \int_{\mathbb{R}+^4}  d \epsilon_\alpha d \epsilon_\beta d
\epsilon_j d \epsilon_k \frac{\delta (\epsilon_\alpha + \epsilon_\beta +
\epsilon_j + \epsilon_k -eV ) L^2   \Avg{\psi_\alpha(\mathbf{x}_S)
  \psi^*_\alpha (\mathbf{x}_S)    \psi_\beta(\mathbf{x}_D)
  \psi^*_\beta (\mathbf{x}_D) }   }{ (E_C +\epsilon_\alpha +\epsilon_k
)^3 } \label{eq:in-misura-in}\\
\int_0^\infty ds\, \mathcal{F}_\textrm{el} (s) =& - i e \,
\frac{G^{(S)} G^{(D)}}{4 \pi e^4} \, eV
\, \int_{\mathbb{R}_+^2} d
\epsilon_\alpha d \epsilon_\beta 
\frac{L^2
  \Avg{\psi_\alpha(\mathbf{x}_S)   \psi^*_\alpha (\mathbf{x}_D) 
  \psi_\beta(\mathbf{x}_D)   \psi^*_\beta (\mathbf{x}_S) }}{ (E_{C}
  +\epsilon_\beta) (E_{C}
  +\epsilon_\alpha )^2} . \label{elastico-misura1}
\end{align}
Indeed Eqs.~(\ref{eq:in-misura-in},\ref{elastico-misura1}) show that
$\int_0^\infty ds \, \mathcal{F}_{\textrm{in(el)} } = -
(+) \partial_{E_C} I_{\textrm{in(el)} }$. 
This directly gives the relation between the cotunneling time and the
cotunneling current in Eq.~(\ref{eq:definizione}).

\subsection{Rules for Feynman diagrams}
\label{feynmann}

As discussed in this appendix the correlation function $\Avg{ I(t)
  N(t-s)}$ and the cotunneling 
  current $\Avg{I}$ are calculated perturbatively in $H_T$. 
They are obtained to fourth order in perturbation theory.
The various contributions are expressed in terms of Feynman diagrams. 
We present here the rules to obtain all the diagrams for the
correlation function and their corresponding analytical expression.
The diagrams (Fig.~2
, Fig.~\ref{all-I}
, Fig.~\ref{fig:I-to-IN}
) are drawn on the Keldysh contour (cf.\ \fig{fig:regole}(a)
). 
\begin{figure*}[h]
\includegraphics[width=140mm]{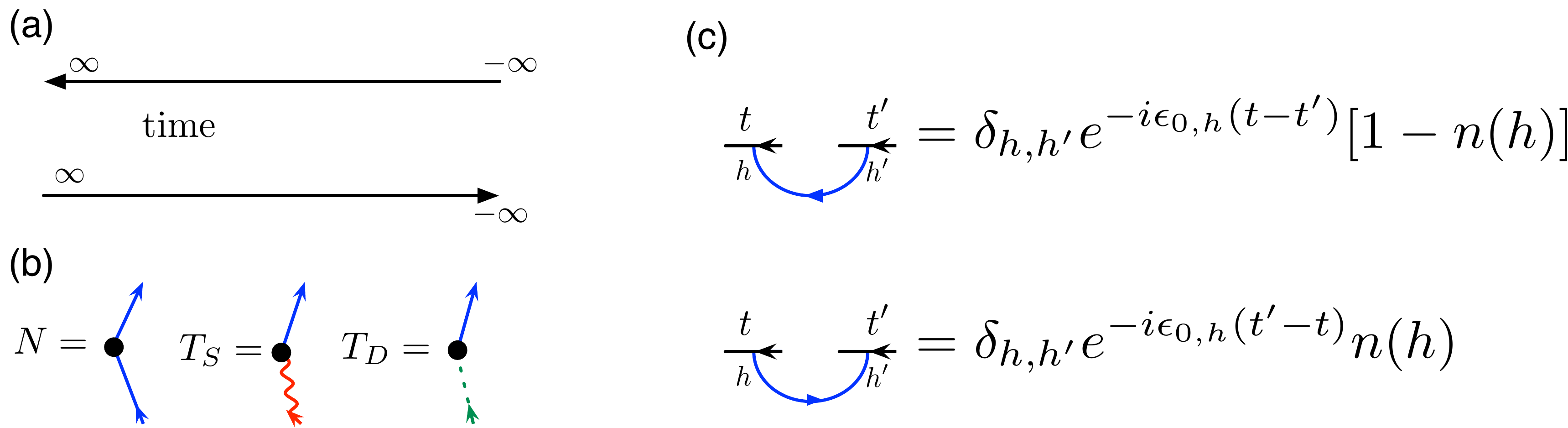}
\caption{(A) Time contour for the Keldysh formalism. (B) Rules for
drawing vertices in the diagrams for $\Avg{I}$ and
$\Avg{I(t)N(t-s)}$, required by the rule {\bf R2}. (C) Explicit expressions
of propagators stipulated by the rule {\bf R8}. 
}
\label{fig:regole}
\end{figure*}
The Feynman rules are:
{\bf R1}. Each operator is drawn as a vertex on the Keldysh contour
according to the expressions given in~\fig{fig:regole}(b)
; the vertices for  $T_S^{\dag}$, $T_D^{\dag}$ are obtained by reversing the arrows
in $T_S$ and $T_D$.
{\bf R2}.  Each operator (vertex) is labeled by a time, and a
subscript ($+$ or $-$)  indicating whether the operator appears in forward- or backward-in-time branch of the Keldysh contour.
{\bf R3}. To begin, the diagrams for the current $\Avg{I}$  are drawn by
inserting the operator ${T_D}_-(t)$ or ${T_D}_-^\dag (t)$; the
diagrams for the correlator  $\Avg{ I(t) N(t-s) }$ require instead
${T_D}_-(t)$ (or ${T_D}_-^\dag (t)$) and  $N_+(t-s)$. 
{\bf R4}. All possible combinations of 
  ${T_S}_{\pm}^\dag(s_2)$, ${T_D}_{\pm}^\dag(s_3)$ should be inserted,
  such that the inequality $s_3 > s_2 >s_1$ is satisfied by the
  operators appearing in the same branch.
{\bf R5}. Vertices should be connected in all possible ways
through the appropriate  propagators.
{\bf R6}. Each time interval between two subsequent vertices is labeled
  by the corresponding charging energy, $\avg{U}$ (cf.\ Section~\ref{modello}). 
This charging energy is set to $0$ at $t=-\infty$.
  Following the time contour, each vertex changes the value of
  this energy in a well defined definite way. 
We label the charging energy at a certain time in terms of
 the vertices that precede that time on the contour. 
We thus introduce $\Avg{U(T_S)}=E_C$ and
 $\Avg{U(T_D^\dag)}=E_C'$ (cf.\ \fig{fig:regole}
). 
All the other charging energies are determined in  terms of these two values. 
In particular $\Avg{U(T_D^\dag T_S)} = -eV$ and $\Avg{U(T_S^\dag T_D)}=  eV$.
An example of energy labeling is shown in the diagram in the inset of
Fig.~2.
.

So far we have listed the rules for drawing and labeling the
diagrams. We complement this list by the rules for \emph{calculating}
these diagrams.

{\bf R7}. Each vertex corresponds to
$T_D \to \sum_{kh} T^{(r)}_{k,h}$, $T_S \to \sum_{kh} T^{(l)}_{k,h}$,
and the respective complex conjugate expressions for $T_S^\dag$,
$T_D^\dag$. 
{\bf R8}. The propagators associated with the dot's dynamics are given by the
expressions in~\fig{fig:regole}(c)
; analogous expressions hold
for the leads' propagators, where the energy $\epsilon_0$ is
replaced by the energy of the modes in the lead, $\epsilon_k^{(\alpha)}$.
{\bf R9}. A factor $e^{-i U (t-t')}$  should be included for the corresponding charging energy  $\Avg{U}$ between times $t$ and $t'$.
{\bf R10}. Integration over times should be executed, accounting for
the inequalities of R4.

\end{widetext}


\end{document}